\documentclass[conference,compsoc]{IEEEtran}
\iftrue
 \newcommand{\mannat}[1]{\textcolor{purple}{\textbf{[Comment: }\textit{#1} -- mannat\textbf{]}}}
    \newcommand{\ha}[1]{\textcolor{red}{\textbf{[Comment: }\textit{#1} -- ha\textbf{]}}}

\else
      \newcommand{\sana}[1]{\ignorespaces}
      \newcommand{\ha}[1]{\ignorespaces}
       \newcommand{\mannat}[1]{\ignorespaces}
\fi

\ifCLASSOPTIONcompsoc
  \usepackage[nocompress]{cite}
\else
  \usepackage{cite}
\fi
\usepackage{enumitem}

\usepackage{booktabs} 
\usepackage{soul}
\usepackage[most]{tcolorbox}
\usepackage{graphicx}
\usepackage{hyperref}
\usepackage{mdframed}

\newtcolorbox[auto counter, number within=section]{summary}[2][]{%
  colframe=blue!80!black, 
  colback=blue!10, 
  coltitle=black, 
  fonttitle=\bfseries, 
  title=Takeaway~\thetcbcounter: #2,#1,
  boxsep=1mm,   
  left=1mm,     
  right=1mm,    
  top=1.3mm,      
  bottom=1mm    
}

\ifCLASSINFOpdf
\else
\fi
\begin{document}
%
\title{Researchers' Perspectives on Navigating the Ethics of Internet Measurement:\\ Case Study of an Internet Measurement Research Group in the EU}

\title{Navigating the Ethics of Internet Measurement:\\ Researchers’ Perspectives from a Case Study in the EU}


\author{
\IEEEauthorblockN{
Sahibzada Farhan Amin\IEEEauthorrefmark{1},
Sana Athar\IEEEauthorrefmark{2},
Anja Feldmann\IEEEauthorrefmark{2}\\[1ex]
Ha Dao\IEEEauthorrefmark{2},
Mannat Kaur\IEEEauthorrefmark{2}
}
\IEEEauthorblockA{\IEEEauthorrefmark{1}Saarland University, s.farhan.amin@hotmail.co.uk}
\IEEEauthorblockA{\IEEEauthorrefmark{2}MPI-INF, \{sathar, anja, hadao, mkaur\}@mpi-inf.mpg.de}
}


\maketitle

\begin{abstract}

Internet measurement research is essential for understanding, improving, and securing Internet infrastructure. 
However, its methods often involve large-scale data collection and user observation, raising complex ethical questions.
While recent research has identified ethical challenges in Internet measurement research and laid out best practices, 
little is known about how researchers actually make ethical decisions in their research practice.
To understand how these practices take shape day-to-day from the perspective of Internet measurement researchers, 
we interviewed 16 researchers from an Internet measurement research group in the EU.
Through thematic analysis, we find that researchers deal with five main ethical challenges: privacy and consent issues, the possibility of unintended harm, balancing transparency with security and accountability, uncertain ethical boundaries, and hurdles in the ethics review process.
Researchers address these by lab testing, rate limiting, setting up clear communication channels, and relying heavily on mentors and colleagues for guidance. 
Researchers express that ethical requirements vary across institutions, jurisdictions and conferences, and ethics review boards often lack the technical knowledge to evaluate Internet measurement research.
We also highlight the \textit{invisible labor} of Internet measurement researchers and describe their ethics practices as \textit{craft knowledge}, both of which are crucial in upholding responsible research practices in the Internet measurement community. 
\end{abstract}

\IEEEpeerreviewmaketitle

\section{Introduction}
Internet measurement helps researchers understand and assess diverse aspects of the web, from user security and privacy to application behavior and traffic patterns~\cite{Ciprian2018}. 
Yet with this exploration come persistent ethical uncertainties,
as behind every packet of traffic is a real person~\cite{Dirksen2024}. 
Researchers who once viewed the Internet as a neutral technical system now see it as a sociotechnical infrastructure where even a simple measurement can affect people and their livelihoods~\cite{sep-ethics-internet-research}.
This shift highlights unresolved tensions: obtaining consent from millions of unaware users is practically impossible~\cite{sep-ethics-internet-research, pauley2023}, and even well-intended experiments can disrupt networks or harm operators~\cite{Fiebig2024}. 
Each new measurement is a balance between the benefits of knowledge and the risks imposed on those who never agreed to be ``subjects'' in the first place~\cite{sep-ethics-internet-research,pauley2023}.

Prior work has highlighted these challenges including the possibility of unintended harm~\cite{Fiebig2024, Florian, Partridge2016, 10.1145/2793013.2793021}, privacy and consent issues~\cite{Huz2015MTurk, Partridge2016, 10.1145/2793013.2793021, Ciprian2018, pauley2023, Florian} and inadequate ethical guidelines and review processes~\cite{Fiebig2024, Florian, Kohno2023, pauley2023, Finn2023Menlo, Zhang2022, Ethics2017, Partridge2016, Huz2015MTurk, Dirksen2024}.
The recommended strategies for best practices emphasize transparency and accountability~\cite{Florian, Fiebig2024, DScope2023, pauley2023, Finn2023Menlo, Zhang2022, Mazel2017, Ciprian2018, Ethics2017, 10.1145/2793013.2793021, Huz2015MTurk}, minimizing harm~\cite{DScope2023, 10.1145/2793013.2793021, Partridge2016, Huz2015MTurk}, and better institutional~\cite{Kohno2023, Zhang2022, Ciprian2018, Ethics2017, 10.1145/2793013.2793021, Partridge2016} and community-level~\cite{10.1145/2793013.2793021, Ethics2017, Kohno2023, Fiebig2024} support.
The Menlo Report~\cite{Menlo2012} expanded research ethics from biomedical studies to information and communication technology (ICT), providing foundational guidance for Internet measurement.
Institutional review boards now scrutinize such projects more carefully, and leading conferences increasingly require ethics sections in submitted papers~\cite{IMC2024,PAM2024,TMA2024}. 
However, many papers simply state that their work “raises no ethical concerns” without deeper reflection~\cite{Dirksen2024}. 
This apparent compliance masks a key gap: we still know little about researchers' perspectives and practices on resolving ethical dilemmas in their everyday work. 
Existing literature on the ethics of Internet measurements does not yet include the lived experiences of Internet measurement researchers.

In this paper, we examine the ethical practices of Internet measurement researchers. 
Focusing on an Internet measurement group in the EU as a case study, we engaged with 16 researchers working on different sub-fields of Internet measurement, using different research methods and with varied research experiences. 
Rather than prescribing universal solutions, we seek to understand how researchers actually make ethical decisions when faced with uncertainty, competing values, and practical constraints. 
We address the following research questions:

\begin{mdframed}
\textbf{RQ1}: 
What are the ethical challenges faced by Internet measurement researchers in their studies and how do they address them?
\end{mdframed} 

Researchers described facing five recurring challenges in Internet measurement studies: privacy and consent issues, the possibility of unintended harm, difficulties with transparency and accountability, uncertainty about ethical boundaries and hurdles in the ethics review process.
To address these dilemmas, participants relied heavily on mentors and colleagues rather than formal procedures.
They adopted practical strategies-also outlined in prior literature-such as lab testing, conservative scan rates, tailored protocols, and transparency measures like reverse DNS, contact information, and rapid responses.
They emphasized however that despite these efforts, harm cannot be fully avoided and decisions are often situational.

\begin{mdframed}
\textbf{RQ2}: 
Do Internet measurement researchers rely on formal ethical processes or guidelines? If yes, which ones, and how do they influence their
decision-making processes? 
\end{mdframed}

Researchers reported that formal ethical frameworks and oversight bodies play only a limited role in their decision-making. Many felt that IRBs lacked the technical expertise to identify hidden risks, often approved projects without fully understanding the implications, and introduced delays or inconsistencies across institutions. While some participants cited the Menlo Report~\cite{Menlo2012} as a baseline, most relied on personal reasoning, tool documentation, or informal community norms when guidelines were unclear. Mentors, supervisors, and ethics sections in top-tier conference papers provided more practical guidance than formal guidelines (such as the Menlo Report), though participants noted that these ethics sections were often inconsistent or performative. Overall, researchers saw frameworks as reference points but found real decision-making shaped more by community practice, peer discussion, and individual judgment.

\noindent Our study makes the following main contributions: 
\begin{itemize}[leftmargin=4mm]
    \item We are the first to address a literature gap by describing the perspectives and experiences of Internet measurement researchers regarding the ethical aspects of their work.
    \item We highlight the unrecognized-often invisible-work performed by Internet measurement researchers that is crucial in conducting responsible Internet measurements.
\end{itemize}

 \section{Related Work} \label{literature}
Internet measurement research has shifted from viewing networks as purely technical systems with little human impact to recognizing them as sociotechnical infrastructures deeply intertwined with human action~\cite{Mazel2017}. 
In addition, Pauley and McDaniel~\cite{pauley2023} show that large-scale Internet measurements inevitably involve human subjects and raise unresolved questions of consent and data protection.
Fiebig~\cite{Fiebig2024} further emphasizes that researchers may cause unintended operational harm, such as service disruptions, system overloads, or crashes, when conducting Internet measurements without sufficient configuration or operational knowledge. 
As a result, ethics has become a central concern in Internet measurement research, shaping how studies are designed, reviewed, and reported.

Researchers in Internet measurement encounter several recurring ethical challenges. First, active probing or scanning can unintentionally cause harm by disrupting services, overloading systems, or placing an excessive burden on network operators~\cite{Fiebig2024,Florian}. 
Second, privacy and consent issues frequently arise, as large-scale data collection often captures sensitive user information without the possibility of obtaining traditional informed consent~\cite{pauley2023,Ethics2017}. 
Third, researchers face legal and regulatory ambiguity, especially in cross-jurisdictional contexts such as server-side scanning and Internet filtering, which can lead to legal consequences~\cite{Florian}. 
In practice, researchers also encounter dilemmas such as limited support from legal departments and the absence of domain-specific guidelines, which hinder consistent ethical treatment~\cite{Zhang2022}. 
The gap between what the law requires and what can be enforced in practice remains one of the most serious challenges in Internet measurement ethics.
Finally, researchers must contend with the limitations of ethical review. Traditional oversight mechanisms such as institutional review boards (IRBs) and ethics committees often lack the technical expertise to assess the indirect human impact, cross-disciplinary complexity, and infrastructural risks of Internet measurement studies~\cite{Ethics2017}. 
Therefore, they may apply inappropriate standards or overlook key issues, leaving researchers without adequate guidance. 
These challenges highlight the need for researchers to continually navigate the tension between advancing scientific knowledge and protecting the interests of stakeholders.

Researchers have proposed several frameworks and guidelines to address the ethical challenges of Internet measurement. 
The Menlo Report~\cite{Menlo2012}, modeled after the Belmont Report~\cite{belmont1979}, is the most widely cited, adapting human-subjects research principles to information and communication technology. 
Scholars frequently reference it as the ethical foundation for responsible measurement~\cite{Fiebig2024,Florian}, yet others highlight gaps between its principles and their application in practice, particularly regarding sensitive data and community norms~\cite{pauley2023,10.1145/2793013.2793021}. 
Beyond such frameworks, researchers also apply broader ethical theories: consequentialist reasoning emphasizes weighing harms and benefits, while deontological approaches stress duties and rights~\cite{Kohno2023}. 
To ease the burden on individual researchers, Dirksen et al. \cite{Dirksen2024}  suggest systemic alternatives such as ``federated ethics boards'' that distribute responsibility across institutions. 
These frameworks and proposals illustrate both the progress and the persistent gaps in guiding ethical Internet measurement research.

While the literature on ethics in Internet measurement research has made significant steps in identifying ethical challenges, developing frameworks, and recommending best practices, a critical gap remains: there is limited understanding of how researchers themselves navigate ethical dilemmas and make ethical decisions in practice. 
Much of the existing literature explores ethical concerns from the perspectives of network operators, ethics reviewers, and legal entities~\cite{Florian}; and of security researchers~\cite{Zhang2022}, but there remains a significant gap in understanding the ethical decision-making processes of researchers themselves.
This gap in the literature highlights the need for further research that examines the experiences of Internet measurement researchers directly. 
By interviewing researchers about their ethical considerations, challenges, and decision-making processes, we gain valuable insights into how ethical principles are applied in practice, what factors influence ethical decisions, and what resources or support researchers require to conduct ethically sound research. Such research would complement existing literature by bridging theory and practice, identifying practical challenges, recognizing contextual factors, and tracking the evolution of ethical norms.

\section{Qualitative Methodology} \label{Methodology}

\subsection{Participants and Recruitment} \label{participants}
For our case study, we engaged with participants from an Internet measurement research group in the EU.
Participants' research areas can be found in Table \ref{tab:comprehensive_background}.
However, we do not share any further demographic details, such as location or the level of academic experience of the participants, to maintain participant anonymity.
Overall, the study included 16 participants representing a broad spectrum of research experience, ranging from master’s students to experienced researchers.

The group performs both active and passive measurements: large-scale Internet-wide scanning, including IPv6 measurements, web-based scanning, SSH investigations, and analyzing traffic patterns, routing data, and tracing routes. 

\begin{table}[!htbp]
    \centering
    \caption{Overview of Participant Research Backgrounds.}
    \resizebox{\linewidth}{!}{

    \begin{tabular}{p{2.5cm}r|p{2.6cm}r}
    \toprule
    \textbf{Research Area} & \textbf{Methods} & \textbf{Research Area} & \textbf{Methods} \\
    \midrule
    Internet Infrastructure & P & Network Security & A/P \\
    \hline
    Routing Security & A/P & Security Protocols & A \\
    \hline
    Network Protocol & A/P & Network Topology & P \\
    \hline
    Transport Layer & Lab & Social and Censorship & P \\
    \hline
    Internet Scanning & P & Security & A/P \\
    \hline
    Internet Censorship & A/P & DNS Infrastructure & A/P \\
    \hline
    Routing Analysis & P/LA & Protocol Deployment & A \\
    \bottomrule
    \end{tabular}
    }

    \vspace{1mm}
    
    {\footnotesize \textbf{Legend:} A = Active Measurements, P = Passive Measurements, 
    \\LA = Limited Active, Lab = Laboratory-based}
    \label{tab:comprehensive_background}
\end{table}

We ensured that the participants were fully informed about the study and participated voluntarily (i.e. no financial compensation). 
We contacted all researchers in the department, and those
interested (16/26) were given the informed consent form (see Appendix \ref{Consent Form}) which described the study’s objectives, procedures, data handling, and participants’ rights including the right to withdraw and data deletion.

\subsection{Interviews} 
Semi-structured interviews were chosen to allow for a natural conversation flow (see Appendix \ref{Interview Structure}).
A pilot study was conducted to test the interview protocol.
The conversations began by discussing the participant's past research experience and current research focus, naturally transitioning to their experiences with regards to navigating the ethical aspects of their work.

Interviews were conducted online on an internally hosted instance of BigBlueButton (BBB)\footnote{Available at - \url{https://bigbluebutton.org/}}.
Online interviews offered flexibility for the participants in scheduling the interviews, and using an internally hosted instance of BBB helped to ensure the security and privacy of participants' data.
The interviews lasted between 45 to 60 minutes.

\subsection{Thematic Analysis}
We opted for an inductive Reflexive Thematic Analysis (RTA) approach~\cite{Braun08082019, Braun03072021} to find patterns and meaning in qualitative data.
An inductive approach is data-driven, meaning that theme formulation relied solely on participants' inputs.
This allowed for a deep engagement with the data, highlighting the participant's nuanced experiences and perspectives~\cite{Braun01012006}.

Phase 1 - \textit{familiarization with the dataset} - involved transcribing the interviews using an internally-hosted instance of the `Whisper' speech-to-text model\footnote{Available at - \url{https://github.com/openai/whisper}}.
Since automatic tools may transcribe incorrectly, we listened to each recording again to fix such errors.
We also removed any Personally Identifiable Information (PII) from the transcripts to maintain participants' anonymity. 

Phase 2 - \textit{coding} - involves labeling qualitative data.
This step was performed manually using ATLAS.ti\footnote{Available at - \url{https://atlasti.com/atlas-ti-desktop}}.
We coded the transcripts to capture both semantic and latent codes \cite{Braun08082019, Braun01012006}. 
This meant labeling both direct answers (semantic) and also underlying beliefs (latent).
This thorough approach led to 245 main codes, which had a number of sub-codes, making it a total of 1,226 codes (see Appendix~\ref{subsec:codebook}). 

We used Miro\footnote{Available at - \url{https://miro.com/app/dashboard/}}, to group similar codes to identify patterns and relationships in the data, and \textit{generate initial themes} in phase 3. 
These groups led to the creation of initial themes.
In phase 4 - \textit{reviewing and refining themes} - we checked that the initial themes showed a clear and separate pattern and were reflected in the participant inputs.
Through continuous refinement, we split broad themes into more specific sub‑themes, merged closely related themes, and reorganized them to better reflect the data.

Phase 5 involved \textit{defining and naming the themes}.
We named the themes to ensure they were concise but also descriptive.
We ended up with two main themes (Section~\ref{theme1} and~\ref{theme2}).
Finally, \textit{producing the report} constitutes phase 6.
For every theme, we described what it means using participants' quotes all the while connecting our findings to prior work (see Section~\ref{findings}).

\paragraph{\textbf{Reflexivity}}
Our multidisciplinary team consists of researchers with expertise in Internet measurement, online privacy and human factors research.
We are motivated to understand and support ethical people-centered Internet measurement research practices.

\section{Findings} \label{findings}
The first theme (Section~\ref{theme1}) answers RQ1 and the second theme (Section~\ref{theme2}) answers RQ2.

\subsection{Ethics Challenges and Coping Mechanisms}
\label{theme1}

Active measurement researchers face privacy and transparency issues, and the risk of causing accidental harm to systems.
Passive measurement researchers, meanwhile, deal with privacy concerns and handling sensitive data that they collect from various sources. 
Security researchers navigate unclear legal rules and complex decisions about whether the benefits of their research work outweigh the possible harms.

\subsubsection{Privacy and Consent Issues} 
\label{Privacy and consent issues}

\textit{``Privacy is a key element''} (P1) in Internet measurement research.
In contrast to studies that involve humans directly, Internet measurements usually gather data from a large number of users without their knowledge.
Modern Internet users freely share data and it is almost unavoidable that sensitive data is captured from Internet measurement which poses serious challenges for research ethics~\cite{soe2023ethics}.
Due to the nature of Internet measurements, it is not possible to take consent from everyone, especially in large-scale measurement~\cite{pauley2023}.
\textit{``Gathering Internet traffic is traffic that usually generated by people''} (P3), even if researchers do not interact with users directly.
This is because the \textit{``Internet is such an interconnected system there is no way around it''} (P8) and there is always a risk of observing personal or sensitive information. 
Everyone can see the data and misconfigurations and \textit{``everyone is affected by it''} (P8). 

Technical data is prone to privacy risk 
\textit{``because there could be some personal data somewhere in there''} (P14).
Researchers admitted that \textit{``it's an unrealistic scenario''} (P10) to get consent from every Internet user affected by their measurements. 

\textit{P10: ``We literally went and scanned the whole Internet. So that would mean I have to contact the 60,000 AS\footnote{Autonomous System} administrators. That would have taken a few years to get all the replies and consents."}

P14 pointed out that the scale of the Internet makes informed consent infeasible \textit{``because there's too many actors involved''} (P14) and they \textit{``have to work on this notion of implied consent''} (P14).
The inclination to study the global Internet while lacking a mechanism to inform or ask everyone involved highlights a fundamental tension.
In specific cases, proactively approaching the network operators might be feasible, however, such direct engagement requires significant resources and is not scalable.

\paragraph{\textbf{Perceived Lower Risk in Passive Measurement}} 
\label{PM and Perceived Lower Risk}
In passive measurements, e.g., observing public routing announcements or aggregate traffic statistics, the expectation of individual consent is mitigated by the public or indirect nature of the data. 
For example, an Internet researcher focusing on routing data insisted that \textit{``since it's public [...] data and we have no immediate humans involved, it's not really privacy sensitive''} (P8).

P5 noted that working in a closed lab-based testbed avoids privacy concerns entirely as \textit{``it doesn't affect other people's network traffic''} (P5).
They view their work as studying systems rather than people. 
Nonetheless, passive data can sometimes contain hidden identifiers or enable inference about users.
Researchers may start with the intent to avoid personal data, but \textit{``the structural nature of the Internet is such that you can't have strict privacy boundaries something personal can always sneak in''} (P4).
Working with public data is \textit{``not really privacy-intense… you see misconfigurations, but everyone sees them''} (P8).
This assumes implicit consent regarding public data sources, essentially relying on an opt-in by default model.

Nevertheless, such assumptions can be problematic; even public data can raise privacy issues if repurposed or aggregated in unique ways. 
Participants recognized this, acknowledging a lingering \textit{``risk of re-identification... despite lack obvious PII''} (P10).

\textit{P10: ``We see the IPs and therefore we can find a network administrator, we can find geo locations.}"

This leads to \textit{``countries and this ends up in politics''} (P10),
meaning that even from public data one can retract information that could be politically sensitive. 
Moreover, there is the risk of misuse of this data if made public, which \textit{``would be a wonderful target list for the bad guys''} (P10).

\paragraph{\textbf{When Metadata becomes Personal Data}} 
\label{Metadata as Personal Data}

When collecting data either actively or passively, it contains some hidden identifiers which may not seem obvious at first.

\textit{P16: ``Sometimes you only later learn like you designed your experiment and during data analysis, you figure out that you actually measured something else along with the data you collected. Sometimes you can have actual content, like something about people just hidden in metadata.''}

Another participant clarified that the metadata -- even when aggregated -- reveals a lot more than pin-pointing the identity of the user per se. 
It might reveal behavioral and social patterns.

\textit{P3: ``it's not necessarily exposing their identity there could be exposing some thoughts and interests that might be different from society [sic] norm.''}

This information when analyzed or published could lead to cultural or communal visibility that could indirectly inflict harm through social judgment or stereotyping.

\textit{P3: ``talking about people with different ethnicities if we're looking at what are they interested in or what are their daily habits they might be a bit different than the overall society that they're living in and just maybe revealing these [...] cultural differences that are unknown to everybody might put them in pressure and I mean people might look at them differently.''}

P4 who does security analysis pointed out that whilst analysing adversary's behavior they can stumble upon sensitive information which are hidden in the metadata but can potentially be PII:

\textit{P4: ``let's say that I get logs about a certain device and [...] this malware is basically something very specific to a VOIP telephone. Let's say right now I kind of know that this is a VOIP telephone so well this is to a certain level identifiable information and now if I look at the IP and where it is geo located I know that this is probably let's say the Middle East. So I know this is a VOIP telephone the Middle East [...] the point is it's very inferable.''} 

\noindent \textbf{Coping with Privacy and Consent Issues} 
\label{coping_privacy}
Below we present the ways in which participants deal with these.

\paragraph{\textbf{Anoymization and Aggregation}}
When potentially sensitive data must be collected, careful handling and de-identification to protect privacy was emphasized.
Many described using anonymization or aggregation techniques before analyzing or sharing data, because it's \textit{``really important to keep the data anonymized''} (P3) and ensure that \textit{``it's not possible to fingerprint a person"} (P3).

Some shared technical and administrative measures help protect the data, e.g. not uploading it \textit{``to any external services"} (P16) and not sharing \textit{``it publicly''} (P16), thereby minimizing third-party risk.
Hashing and anonymizing identifiers (P15) was also mentioned.

\paragraph{\textbf{Balancing Privacy and Data Utility}}

\textit{``Anonymization is almost always at odds with the validity of our research... it destroys the very phenomenon we are seeking to study.''} \cite{pauley2023}.
This tension surfaced in our interviews. 
One researcher explained why they do not anonymize certain data (like IP addresses) during collection:

\textit{P1: ``No, I'm not anonymizing some of the data right away, I'm not able to study these behaviors [...] but I'm also not sharing this data.''}

The researcher chose to keep the sensitive data for analysis so that the accuracy is preserved. 
To prevent privacy breaches, they treat the data as highly confidential.

\paragraph{\textbf{Controlled Access and Non-Disclosure}}
\label{cand}

Good access control mechanism helps in data protection, e.g. storing the data on local servers to minimize any third-party data leaks and also limiting access to who can access the data. 
However, access control also occurs in informal ways, such as in internal communication channels that \textit{``are shared only with the designated persons''} (P10) and not sharing sensitive data \textit{``even for making slides or chatting with colleagues''} (P3).
P16 does not \textit{``upload it to any external services clouds''} but stores \textit{``such data on internal systems''}.
Some noted not disclosing certain datasets at all, even in academic publications.
P1 who collects attack traffic described that the collected data \textit{``is going to have an ID for a compromised host or an attacker's system. So I never share a dataset.''}
These details, if they become publicly available, could inadvertently expose victims (compromised hosts) or even aid malicious actors. 
Others working with \textit{``a commercial dataset''} (P9) from companies similarly pointed out being constrained by confidentiality.

\paragraph{\textbf{Alternatives to Individual Consent}}
 
Researchers often deliberately avoid or minimize the collection of PII right from the beginning of their research design process. 
 
 \textit{P1: ``I'm not interested in how the users behave, so I'm not collecting user data.''}

By structuring their studies around publicly available datasets or already accessible aggregated information, researchers aimed to navigate around the complex requirements for individual consent altogether.

When the user is engaged directly (which is rare in Internet measurement), then the researchers obtained informed consent.
P11 shared that if they \textit{``have volunteers, yes, consent is required''}.
They were conducting a human-subject experiment where the recruited participants are asked to run measurement software.
For this, they followed the established ethics process for human subjects studies.

Another participant described an instance of negotiating permission with an infrastructure provider, underlining
that the consent from organizations, for instance ISPs or cloud providers, may also be needed apart from individual users when the study is dependent on their data or networks.
These negotiations could be time-consuming and unpredictable.

\subsubsection{The Possibility of Unintended Harm} 
\label{Unintended Harm}

\textit{``Any research could lead to harm. It would be foolish if someone said no"} (P1).
From network disruptions to exposing vulnerabilities to even misleading other researchers.
\textit{``It's always a possibility''} which we \textit{``can't fully exclude''} (P15).
Several participants pointed out that active scans are unpredictable. \textit{``There's always a factor that we can miss''} (P1) and \textit{``also a change of behavior that was not expected that impacts the measurement''}.
This uncertainty aligns with observations from prior work~\cite{Kohno2023}.
\textit{``If you do active scans, for example, then you can really break things''} because \textit{``the Internet is held together with like glue and luck and hope. But hope is diminishing''} (P8).

\paragraph{\textbf{Operational Disruption}} 
\label{Operational Disruption}

Service disruptions or system failures can occur. 
\textit{``You might accidentally cause harm''} and \textit{``overload remote systems''} (P16).
\textit{``Sometimes you might send messages to systems for measurements for probing that might crash systems and you didn't even know''} (P16).
The harm depends on \textit{``what disruption level is"} (P2).

\textit{P10: ``There was one router that had linked to it, I don't know, 4 billion IPs or something. So just all of them, /64 on that router [...] When we went and scanned for 1 million IPs, ... this actually, at some moment, just simply overwhelmed that router and no one could connect to it anymore.''}

The disruption occurred because of a device that was misconfigured in the network by the operators, highlighting uncertainty despite careful research consideration.
Sometimes the missteps lie on the researchers' side.

\textit{P8: ``We had occasions where, for example, one of our PhD students [...] did an active measurement. And the first active measurement [...] killed BGP for the whole institute.''}

P8 further added:

\textit{P8: ``...if you're off the Internet for a couple of seconds, that means all of the outside connections are gone. You cannot, you don't have Internet, right? And he did that with an active scan. So God knows what else can happen."}

This shows how poor configurations or intensive scanning can disrupt crucial systems, potentially disrupting the Internet connectivity of the whole organization. 

\textit{P14: ``we are mostly concerned with not overloading the servers, not accidentally causing crashes, and not putting some burden on other people's machines"}

This sense of responsibility to avoid unintended harm is also highlighted in the literature~\cite{Florian, Huz2015MTurk}.

\paragraph{\textbf{Collaboration with Adversaries}}

P1 who works on security analysis explained the ethical dilemma that \textit{``in a way you are contributing to the attacker's topology''}, effectively becoming part of an attack chain or a botnet and \textit{``collaborating with the attackers somehow''}.

This raises the question: at what point does observing merges into enabling harm? 
P1 described this as \textit{``another perspective of ethics''} that they had to consider, which made them either halt their research or alter the experiment in order to avoid unintentionally facilitating criminal activity. 

P1 whilst working on a compromisable system unexpectedly found that \textit{``somehow the system started to receive user data''.}
Little did they know that \textit{``this compromisable system can be exploited in a bigger scheme''}.
As a result, their test system was co-opted by a botnet which caused traffic from unsuspecting user hosts to flow into their experiment.
This put them into an unexpected scenario where they collected the data from the actual users inadvertently who actually had no idea that their compromised machines were sending traffic via the researchers' server.
Ultimately, they chose to discard the (potentially interesting) data only to prioritize user protection over research outputs.

\noindent \textbf{Coping with Unintended Harm}
\label{coping mechanism unintended harm}

Here we present how the participants deal with the possibility of unintended harm.

\paragraph{\textbf{Lab Testing and Staging}}

\textit{``First of all, test your software, everything in lab''} (P10).
P10 explained:

\textit{P10: ``We make a small in lab experiment and try everything that comes to your mind. Ask your colleagues, try those things as well. Once that is set up, ...get yourself someone who actually runs something on the Internet, try your setup against their network. And see how your tools are working in the wild and see if you actually bother that person when you do this''.}

P14 shared that they \textit{``first try to do some small scale experiments to see if your idea works, your idea scales''} (P14).
\textit{``Try to test things in a local environment} (P9) before \textit{``doing a large scale measurement''} (P9) and \textit{``account for the load that you will cause on each network''} (P9).

\paragraph{\textbf{Rate Limiting}}

Almost all active measurement researchers apply rate limits or query limits as a way of not overburdening the system or network. 
P15 noted the \textit{``need to find the sweet spot for the scanning rate''}.
P15 also mentioned that they follow general scanning recommendations based on the Menlo Report~\cite{Mazel2017} and also the established community norms, including in prior studies.  
When it comes to choosing the scanning rate, P15 chose a rate lower than that specified by Durumeric et al.~\cite{ZMAP}.

P8 described this setting of rate limit as \textit{``a rule by thumb''} which helped them find the \textit{``sweet spot''} of scanning rate:

\textit{P8: ``So what we basically did is after every attempt,  we just tried with different, just time intervals....we came up with a number... I think it wasn't even one second. It was like for computer time, a really long time, so to say.''}

Even using potentially lower-rates might not be acceptable to operators:

\textit{P8: ``we [...] pasted like one prefix per second or something like that. So it's really not that big of a use. And even that actually got us some results or not results, some people saying, hey, please go slower.''}

\paragraph{\textbf{Communication and De-escalation}}
Transparency acts as a safety measure and also addresses the consent issue (see Section~\ref{proactive transparency}). 
Operators who receive well‑labeled traffic can identify its source and request a halt.
P10 included reverse lookup queries, a website, and an email address as forms of communication to reduce potential misinterpretation of the study and minimize harm.
They also mentioned that system administrators get \textit{``really, really emotional''} (P10) if the scanning is not done carefully. 

\textit{P10: ``de-escalation and explanation and, of course, taking actions like immediately stop scanning that network and so forth, solves the issue.}"

Similarly, P15 mentioned that they respond swiftly to the complaints. 
P16, similar to P10, sets up websites to inform the operators about the scanning and to minimize the harm. 
Due to these communication channels, researchers can swiftly stop the measurement id needed.
However, some operators lack the expertise or resources to respond, limiting the effectiveness of opt‑out systems: 

\textit{P16: ``Operators sometimes don’t even know what scanning traffic is or how to block it; they reboot their systems and never report anything.''}

\paragraph{\textbf{Deciding to Stop}}
Sometimes the only way to avoid harm is to cease an experiment. 
P10 did so after crashing a network router.
P6 uses a low insertion rate to avoid bias, but would stop injecting dummy domains if it affected other measurements. 
P11 argues that when a point comes where the risks outweighs the potential benefit, then they stop the research altogether.

\textit{P11: ``there definitely is points where we say like we cannot do this research in a way that doesn't put people at too high a risk so we cannot do it.''}
 
P16 acknowledged that they occasionally stop scanning high‑risk targets, although they grappled with situations where not scanning might leave vulnerabilities unaddressed. 
Deciding when to stop is a matter of judgment.

\paragraph{\textbf{Balancing Risk Versus Scientific Benefit}} 
\label{balancing}
Unintended harm can be weighed against the potential benefits of measurement. 
P12 emphasized that scanning misconfigurations helps secure the Internet:

\textit{P12: ``I don't see a major concern because most of these innovations are aimed to protect the network and also to improve the security and privacy. I think I look mostly positive, but if there is an obvious flaw in the innovation that could lead to potential harm, we should stop it.''}

P10 similarly argued that scanning helps reveal compromised machines which ultimately helps the operators to take necessary steps to mitigate the issue or to remediate. 
P15 emphasized the benefits of their work:

\textit{P15: ``we have to scan to some degree to get this information on the network [...] All those studies in this area helped improve the network by now, improve the overall functionality and security for everybody. So I think that it's worth doing it.''}

Such balancing of consequentialist vs. deontological perspectives is reflected in prior work~\cite{Kohno2023}.
P14 feels that \textit{``you can't really assess the probability of this risk. So you can't really balance it, in my opinion''}.
P11 notes that there is a risk involved when conducting certain types of research but mentions weighing \textit{``potential risk and potential use case''}.
Research on censorship circumvention serves as an example where the societal benefits could outweigh its risks.

P9 prefers someone else to evaluate the risk-benefit:

\textit{P9: ``it would be harder for me to judge that because if I'm interested in the topic, then I probably will have a more biased view toward it. [...] So I think it will be better to have someone else evaluate that.}"

On a different note, P3 argues that \textit{``if we always want to be so conscious and just try not to make any harm at all, we might not do any research''} (P3).
Despite this, P3 also avoids measuring sensitive demographics to prevent harm to vulnerable population. 
Similarly, P11 stops the research altogether when the risk is too high. 
However, for several researchers, the importance to conduct new research for safety and user protection outweighs the risk of causing temporary disruption.
Nonetheless, many remain cautious.
For instance, P1 argues against conducting the research \textit{``if you don't have the balance''}.
P10 follows `win some, lose some' approach.
P10 also argues that in theory, ethical review boards should be the ones who would ultimately decide if the benefits of the research outweighs the risk or vice versa. 
P16 summed it all up:

\textit{P16: ``you can't say, There's this golden arrow, which you can just shoot, and then everything's fine.''}

They further added that:

\textit{P16: ``there's no blanket answer for that.  In almost every single case, you have to do that assessment, do a harm-benefit analysis,
 think about what might or might not be going wrong, and then react to that for that specific study.''}

The variation suggests that researchers calibrate their threshold for acceptable harm based on their disciplinary norms, personal ethics, and research experience.

\paragraph{\textbf{Institutional and Community Responses}}

Similar to prior work~\cite{Fiebig2024}, our participants bring up community infrastructure and collaborative approaches.
P15 advocated for collaboration to reduce duplicated scans which will reduce the overall potential for harm as well as the network or system load. 

\textit{P15: ``So we collaborated with several other groups doing similar research, if they already have scans, or if they already have the data, we don't have to do it ourself, to add additional traffic to the network.''}

P6 emphasized the importance of standardized guidelines for scanning rates, contact information and opt-out protocols. 
P16 calls for distributed ethical review with technical experts who understand Internet measurement, rather than leaving decisions solely to researchers or ethical boards. 

\subsubsection{Difficult to be Transparent and Accountable}
\label{difficulttransparent}
Transparency and accountability emerged as critical challenges, particularly for active measurements.
Making research activities discoverable and comprehensible to potentially affected parties while maintaining the validity and efficiency of the research is \textit{``quite challenging'}' (P3) because of the need to ensure \textit{``that the data remains anonymized''} (P3). 
However, transparency is not always possible:

\textit{P4: ``I have done DPI\footnote{Deep Packet Inspection} in the past it's not very transparent in the first place.''} 

\paragraph{\textbf{Competence, Awareness and Frustrations}}
\label{technical competence and discovery}
Transparency mechanisms assume a level of technical competence that all network operators may not possess.

\textit{P16: ``the first big challenge is learning whether something actually happened. Because usually people don't tell you, even if you follow all the good practices of Internet measurement... follow all the little nooks and crannies to get people informed that hey something crashed. Sometimes they don't even know it was you [...] They might just say, ``Hey, my mail server stopped responding. Let's reboot it." And then it happens again, and they're like, ``Let's reboot it."  And they don't know why that happens. They might even lack the ability to figure out why it has happened."}

P16 further observed:

\textit{P16: ``You will never even get an email to reply to, from those people who are not able to identify that it's you doing something. Not because you're not following best practices, but just because they can't.}"

This observation highlights a fundamental mismatch of technical capabilities,
also shared by P10:

\textit{P10: ``there was one case when one system administrator was really angry because of this. But of course, that system administrator was not aware of the Menlo report and all of this, let's say, regulation set up for academic scanning and studies.''}

This unawareness may cause the operators to mistake research for malicious activity which then leads them to \textit{``blocking their entire university's IP range''} (P6).
The resolution of such conflicts often requires knowledge of research practices and community norms. 

\textit{P10: ``So there was a really, really long discussion about this and about explaining how this [Menlo] report work [sic]. And then they understood.}"

This educational burden represents a significant but often unacknowledged cost of maintaining transparency in Internet measurement research, which can be understood if you \textit{``just put yourself in their\footnote{the network operator's} shoes”} (P10).

Researchers mentioned the difficulties in communicating with the operators (P6) who can sometimes get angry and frustrated (P16).
In an instance where measurement studies crashed mail servers, operators \textit{``became a bit abusive towards the researchers''} (P16).
Prior work has noted that system operators often work in fast-paced high-pressure environments where they are responsible for uninterrupted system operations, have to quickly resolve security related issues and also support end-users~\cite{kaur2022needed}.
They are often overworked and overburdened, while at the same time their work remains invisible~\cite{Kaur23}.
Such work dynamics can lead to frustrations, specially when an external entity starts to interact with your infrastructure without your knowledge.

\paragraph{\textbf{Documentation and Reproducibility Challenges}}
Researcher accountability involves publishing measurement results. 
Transparency for affected parties vs. scientific transparency creates challenges. 

\textit{P12: ``We provide enough detail about how, for example, data was collected, processed, and used from resources for how long. We also keep it for some time if there is a question or something.}"

However, the sensitivity of Internet measurement data often hinders research reproducibility.  

\textit{P1: ``I never have an artifact published because most of the data that I work with would be considered with some type of sensitivity."}

Transparency requirements conflict with security considerations (also discussed in Sections~\ref{cand} and~\ref{Unintended Harm}). 
This limitation creates challenges for scientific peer review and replication while protecting privacy and security.
Regarding transparency of ethics, many participants include ethics sections in their papers.
However, P16 criticizes the ethics sections in research papers as being perfunctory: 

\textit{P16: ``these ethical considerations became kind of a fig leaf. Like we did the ethics dance if something went wrong it's not our fault.” }

P14 noted that some venues require only a brief ethics statement, and enforcement is inconsistent. 
P15 believes authors should explain their ethical reasoning in detail rather than providing a single sentence. 

\textit{P15: ``we openly discuss which ethical measures we did or how we configured our scans. I think the trend towards that direction is there, it just has to go further over the next years.''}

\noindent \textbf{Coping with Transparency and Accountability Challenges}. Below we describe how researchers handle these.

\paragraph{\textbf{Proactive Transparency}} 
\label{proactive transparency}

Researchers often proactively inform stakeholders about their scanning activities, usually setting-up an informative webpage. 
Several mentioned maintaining a reverse DNS entry or WHOIS record for their setup with their contact details.

\textit{P9: ``So one of those is to host a website, for example, that contains all the contact information that is needed to opt out on the scanning machine itself. So when you see traffic coming to you from a specific IP address, you can go to that. You can look up the IP address or visit the website that is hosted on that IP address and then find all the contact information that you need. We can have a descriptive WHOIS entry in the database and the Internet registry. Provide all the details for people that want to reach out and opt out of the measurement so they can use that contact. If people happen to actually reach out and complain or even try to opt out, be open to describe what you are doing. So provide some details about the project, why this is important and what exactly are you trying to achieve.''}

This transparency extends to publishing:

\textit{P8: ``we published a big data set of everything that we did. So everyone has the raw data, past data and the exact setup, what we are doing. And people can also ask us about … the scripts that we basically used for all of this. And people actually asked me, so I really, I gave it to them. Why wouldn't I? [...] we try to be as transparent as possible.”}

\paragraph{\textbf{Abuse Handling and Response Mechanisms}}
While institutional support helps in achieving transparency and accountability, their effectiveness is highly dependent on individual researchers' implementation and responsiveness.

\textit{P10: ``we tend to answer in as soon as possible. So in 24 hours, we usually would answer all of this. If you send an email [...] up to 24 hours, there will be answer. Usually it's a few hours, but of course, if it's overnight, there might be more than, let's say, five, six, seven hours until we answer.''}

P15 similarly mentioned that the reaction time should be kept to minimum accompanied with continuous monitoring of \textit{``those abuse email contacts also on the weekend on public holidays over Christmas, whatnot''}.

\textit{P2: ``One thing that I learned is that don't start your measurement or scanning on the weekend. Because people [...] they don't answer. If something broke.”}

The challenge amplifies when research spans multiple institutions or involves automated systems that may not have dedicated personnel monitoring communications. 
This distributed and automated nature of modern Internet measurement complicates traditional accountability mechanisms.
In case of an event \textit{``de-escalation and explanation and, of course, taking actions like immediately stop scanning that network and so forth, solves the issue''} (P10).
This rapid response capability requires dedicated personnel, clear processes and sometimes support infrastructure requiring institutional commitment and resources, which a smaller research group might lack. 
P7 described another approach: 

\textit{P7: ``RIPE\footnote{\url{https://www.ripe.net/}} already follows all the required guidelines and principles and it also has an opt-out feature if you want to opt out of it." }

This reliance on third-party infrastructure simplifies implementation but may limit researchers' ability to customize transparency approaches.
Participants noted significant variation in the quality and response time of abuse handling across the research community. 
These best practices are \textit{``not written down anywhere''} (P4).

\textit{P4: ``but most of the scanners who scan on the Internet they are anonymous scanners in a way that their IPs don't really have a proper WHOIS record like a completely filled out with proper abuse mails and proper contact details where you could contact the person scanning and tell them that you are not happy with their scanning.''}

This variation in abuse handling and response mechanisms may weaken the overall community trust because it can ultimately cast doubts on the legitimacy of the research.
Researchers in our study felt accountable to report any mishaps but the disclosure \textit{``varies based on the severity of the issue''} (P10).

\textit{P2: ``If I break some other company's network, I just have to say, ``apologies. From my deep heart, I apologize'', because I can't do anything.''}

\subsubsection{Uncertainty About Ethical Boundaries}
\label{uncertainethics}
Ethical discussions in Internet measurement research remain, in practice, highly contingent and subjective - leaving researchers with significant uncertainty regarding the boundaries for ethical Internet measurement.

\paragraph{\textbf{Lack of Clear Guidelines}} \label{lack of guidelines}

\textit{``In a lot of the cases you don't have a guideline''} (P1) and ethics standards may vary across countries.
Researchers consult \textit{``not from one specific source''} (P3) but \textit{``some tips that [are] mentioned in some papers or related works''} (P3).
They also relied \textit{``on common sense as researchers''} (P6).
P12 shared that they usually follow the guidelines, however, there are cases where they are not fully aware of how to interpret them which leads to uncertainty.
Scanning was characterized as dicey and a gray area, where guidelines are mere recommendations and \textit{``there is really no way you can enforce it”} (P4).
In contrast, scanning is also referred to as being \textit{``best practice basically''} (P4).

\textit{P4: ``If people are doing scanning they would to a certain level be reading all the research content that happens there and most research content have a very strong ethical section which talk about …what researchers are doing.”}

While conferences have \textit{``kind of mandated to have an ethical section"}, these can  
lack the depth and act as \textit{``fig leafs"} (P16). 
However, \textit{``people have to reflect''} (P16) on the implications of their own work and that, when done right, can be evident from these ethics sections in the papers. 
However, as P16 said:

\textit{P16: ``If you didn't do it and just write some general blah blah, which also a lot of people do, and the reviewers notice, you usually get guided by the reviewers to write some blah blah blah which actually makes sense. So it's like look we're doing something but it's like it doesn't feel like like deep change within the community recognizing that this is something important they have to do. It's more like a performative action. Like we did the ethics dance if something went wrong it's not our fault.”}

\paragraph{\textbf{Reactive vs Proactive Ethics}}
The aforementioned uncertainties can lead to reactive ethics.
Usually, the reviewing of ethical implications are done \textit{``ex-post"} (P16) and makes the whole ethics as a \textit{``performative check"} (P16) since the \textit{``harm is already done''} (P16).

In relation to conference ethics requirements, P16 said:

\textit{P16: ``It's obviously like a reaction to some also recent high profile cases like this. What was it, Michigan, I think, where they send false patches to the Linux kernel to introduce vulnerabilities, which I mean, other obvious unethical thing. So like this whole system setup is kind of weird. There's also no transparency [...] It's just like you show up with work at a conference, you submit it and well, let's hope for the best.”}

Although \textit{``the big problem with Internet measurements, especially active Internet measurements, is that it's really hard to figure out what might actually go wrong"} (P16).

\paragraph{\textbf{Legal != Ethical}}
\label{legalethical}

Researchers who are new to the field often have \textit{``no experience with these kind of ethics''} (P14) and learn through trial and error and past experiences:

\textit{P14: ``in my previous career... you had to sometimes handle personal identifiable information… this was more of a legal question than an ethics question… we did it because there was the law.”}

However, ethics and legalities \textit{``tend to be two different things''} (P16), and legal boundaries are not always unambiguous either.
Laws governing Internet measurements were generally not written with benign research scans in mind, and they vary widely across jurisdictions. 
Existing literature underscores this profound legal uncertainty~\cite{Florian}. 
Some participants were not clear about the legal implications of their work, some delegated the responsibility to the third-party data providers and some assumed that they were complying with legal regulations. 
However, something permissible in one jurisdiction might be illegal in another, creating a \textit{``very dicey''} (P8) situation. 

\textit{P16: ``in the machine learning community, couple of years ago, there was this big outcry about papers [...] motivated with detecting Uyghurs... Well, if you put this in front of a Chinese IRB, it's more like, no, no, this is a document to protect society, like different perspectives."}

Some described operating under the assumption of \textit{research exception} or academic freedom, but with significant uncertainty about the scope and limitations of such protections.
P10 explained: 

\textit{P10: ``As long as the service is publicly accessible, I guess we are allowed to access it and then scan is just an access there as well.''}

One researcher shares that \textit{``usually in the country where Internet censorship is present, such as China or Russia or Saudi Arabia, it is by law. So, if you're circumventing this or looking at the censorship, how it works, [...] it's illegal. So, you are technically breaking the law in that country."}
Legitimate research can lie in a legal gray area.
The overall responsibility then lies on the researchers who must decide on how to get on with this situation, from where and whom to take advice, and how to publish the findings of their work without putting anyone in legal jeopardy.

A researcher scanning \textit{``the full IP address space''} inadvertently (P15) targeted a \textit{``specific command and control service''} (P15) that was monitored by law enforcement.
This led to questions which escalated to involving \textit{``some other parts of the university''} (P15).
Since the researcher informed the institute in advance about the research, the whole process was less cumbersome and 
the incident was de-escalated with prompt and open communication.

Another participant described a similar issue when conducting active network scans on off-ports which \textit{``triggered so many annoyed people''} (P10).
They further explained the frustration, anger and shock of the concerned party as well:

\textit{P10: ````How could you find us? We were on a hidden server. This was something hidden over the Internet....How do you know about this?''...And we would keep explaining them and they would get angry to anyone. ``Oh, how the censors know? Nobody's supposed to know, this is illegal''."}

The affected party felt that the researcher exposed some sensitive information that was supposed to be hidden.
The explanation by the researcher that everything they did was indeed legal was in vain.
Even when the researcher is doing everything legally and by the book, it can be perceived as illegal and cause distress. 

\noindent \textbf{Coping with Ethical Uncertainties}. Below we describe how researchers deal with these.

\paragraph{\textbf{Consultation and Mentorship}} 
\label{Mentors}
We find a strong reliance on research supervisors or experienced colleagues for ethical decisions involving data sharing (P2), rate-limiting (P2), risk-benefit analysis (P3).
When asked about handling of unexpected ethical issues:

\textit{P7: ``I would consult my supervisor and ask him if what should be how should we proceed with this?''}
 
While seeking advice is good,
it also highlights a gap in formal ethical training especially in the context of peculiarities of network experiments. 
In such an apprenticeship model where students absorb the ethical stance of their lab or supervisor, one needs \textit{``to be careful''} (P13) of being \textit{``influenced by my supervisors' views on ethics''} (P13). 
There creates the risk of an ethical monoculture if everyone defaults to one leader's opinions.
Some researchers also consult with their peers to help identify risks as \textit{``they might have different perspective of harms or consequences''} (P3), going the extra mile to ensure that their work is ethically sound and does not cause any harm. 
Some \textit{``not too familiar with what the legal situation is''} (P6) rely on their supervisors' advice to tackle the legal challenges or to interpret laws.
Researchers \textit{``weigh best practices''} (P11) mentioned in the published studies and consult with the supervisors to \textit{``try to minimize the risk''} (P11).

\paragraph{\textbf{Ultimate Responsibility}}

A sense of isolation in ethical decision-making was noted. 

\textit{P10: ``You, as the person [...] doing the scans, you as a researcher, this is not someone else's job. This is your job.”}

While ethics review boards \textit{should} be able to resolve researchers’ uncertainties, they lack domain-specific knowledge so the responsibility ultimately falls on the researcher. 

\textit{P10: ``More [sic] of the time, they are doctors. And doctors see things differently. They basically ask themselves, ``Oh, is anyone dying during this process?" No, then it should be fine.”}

Researchers \textit{``do not have access to a capable IRB''} (P16) always.
Hence, regardless of ethics approval, one still must be considerate of the potential harms and make the call because \textit{``in the end it's your fault [...] it's always you who does it''} (P16).
Researchers suggested mixed expert panels or consulting domain experts - similar to prior work~\cite{Fiebig2024} - could help both ethicists and researchers make decisions.

\subsubsection{Hurdles in the Ethics Review Process}

\paragraph{\textbf{Time-Consuming}} 
\label{Ethical Approval Takes Time}

Formal ethics review processes \textit{``tend to be a bit slow''} (P16) and can take \textit{``quite a long time''} (P11).
Therefore it helps to talk to the ethics board \textit{``very early''} (P11).
Time constraints also influence the \textit{``kind of research you can do in your project''} (P11).
Researchers may require multiple reviews for one project:

\textit{P10: ``We got three ERB submissions on this project. We started with the initial one, which allowed us [...] four months, then another six months, because we started, we had a lot of demo situation and so forth, and then we got another six months, which scanned over the Internet, and then now we have a continuous scanning allowance."}

Researchers may need to obtain ethical approval before fully knowing what their research might find or what ethical issues might arise during data collection.

\paragraph{\textbf{Limited Domain Expertise}} \label{lack of expertise}

Ethics review boards often lack domain-specific expertise~\cite{Fiebig2024}.

\textit{P15: ``most ERB boards I know don't have the knowledge to check this. So the one in other institute was created for the medicine faculty or for anything that involves actual humans. There was no computer science specialist on the team. So whatever we told them, [...] they have no idea what we were talking about."}

And while having domain-specific expertise in the ethics committee would help, it is not a  blanket solution.

\textit{P14: ``I think if I was being into measurement research and have trouble judging that, I'm not sure how the [ethics board] would judge it."}

Recent work has noted that measuring Internet connectivity might actually bypass government censorship~\cite{wendzel2025survey}, which could put people in legal trouble or put their safety at risk. Without knowing these hidden details, review boards cannot properly check whether the study is ethically sound or not. 
Simple technical descriptions may hide deep ethical issues. 
P16 referenced a study \cite{Encore} that used people's browsers to check for censorship by trying to access blocked websites and has become a key example in talks about the limits of IRB reviews in computer science research especially in Internet measurement.

\textit{P16: ``it's like the classical example of an IRB being like, oh, you're not showing angry colors to people, then you're fine. Well, they might have made people in countries where there might be kind of a little bit of the death penalty on certain site visits, visits those sites, which has very obvious ethical implications, but the IRB was like, ``nope, nope, looks good to me''. Which is a very big problem, right?"}

The review board focused only on typical psychological research methods, not considering the dangers to people living in repressive governments.
However, institutional ethics boards' purpose might not be to protect the people:

\textit{P16: ``I have never seen one of those boards actually take a strong stance on ethics. They are more like institutionalized function to make things go away, like not the papers, but the concerns."}

\subsection{Institutionalization and Practice of Ethics}
\label{theme2}
\subsubsection{Frameworks and Guidelines}

The Menlo Report \cite{Menlo2012} is \textit{``rather old''} (P10) but is the \textit{``baseline thing''} (P16).
It \textit{``tells you, okay, you have to clearly state who you are, what you're doing''} (P10), e.g., having \textit{``a website that if you're gonna look and see who this IP is, there's a website explaining what we're doing and why we are doing''} (P10).

Some use \textit{``the Menlo and Belmont reports usually for ethical considerations''} (P11).
\textit{``There's also a paper, especially for the Internet measurement crowd, on ethical data sharing"} (P14).
In addition, some rely on \textit{``community standards from related work''} (P15).

\paragraph{\textbf{The Power of Community Wisdom}}

Community wisdom is passed via conference papers, professional connections, and shared experiences that build up rules that aren't written down but are still widely known and followed.

\textit{P9: ``there is a like community best practices [...] to set up your infrastructure in such a way that you will be reachable and people can opt out of your measurement."}

The community also learns from successful examples. 

\textit{P11: ``you should look at what the best papers are doing in terms of ethics, because some top-tier conferences have highly strict ethical requirements, and if you follow their best practices, you're going to do mostly fine."}

These practices spread through research papers: 

\textit{P4: ``Because if you are in the community [...] are doing scanning they would to a certain level be reading all the research content that happens there and most research content have a very strong ethical section." }

This creates a cycle where published research both shows what the community thinks is right and also helps shape what is considered ethical. 
However, this community-based approach has some limitations. 
 
\textit{P15: ``It's overall a weird mix of what people do. So there are some community standards that people follow in theory. I'm not sure whether all follow those standards." }

\paragraph{\textbf{Technical Tools as Ethical Teachers}}

Researchers may use a tool \textit{``because it's very popular on the Internet''} (P2) and has \textit{``a very well-written guideline on their website and tool''} (P2).
The popularity and documentation clarity affect how widely it is used, which in turn shapes what is considered ethical in the research community.

\textit{P2: ``Even in their paper, they explain it. So even if some people have a question, they can easily refer their paper to read them and find their questions."}

Technical standards often serve as ethical guidelines as well.
For example, the robots.txt protocol is both a technical rule and a way to set ethical limits.

\textit{P8: ``It basically says you are allowed to automatically access this, that, and that, but you're not allowed to automatically access this, that, and that."}

\paragraph{\textbf{The Mentorship Alternative}}

In place of formal structures, researchers learn through knowledge shared by their academic teachers (also discussed in Section~\ref{uncertainethics}). 

\textit{P1: ``I don't have a book that I'm following this framework that was written down, how should I approach this? I learned from really good people."}

Ethical knowledge is a skill that is passed on through personal connections and shared experiences.

\textit{P12: ``what is more beneficial is the experience of seniors in the group that they bring actively in the project. So when they see it like professors, postdocs, because they have gone through it and they know this might be sensitive information that we should not publish."}

The mentorship approach is also useful during ongoing research when \textit{``there could be cases where you are not aware that this could be interpreted differently''} (P12) or when \textit{``such a situation did not happen''} (P7) before.
However, the mentorship model can create dependencies by relying heavily mentors who are available and competent. 

\paragraph{\textbf{Institutional Patchwork}}

Researchers navigate different rules, different understandings, and different ways in which things are actually done. 

\textit{P12: ``conferences, they have their own policies that need to be considered when you submit your paper. For example, for anonymization of data or for explicitly writing if your work has any ethical impact in your paper."}

The variation in institutional approaches created challenges for researchers working across multiple institutions or collaborating with international partners (see section \ref{legalethical}).
Despite these challenges, participants acknowledged that institutional oversight has value, even when imperfect. 

\textit{P10: ``I was feeling rather safe when doing the scans, given the fact that I was following all the recommendations in the report...I was feeling that I did my part and I was doing everything legally, right? In the proper way, I was not breaking any laws or things like that."}

Institutional guidelines might be good enough for approval processes, but do not provide detailed guidance for day-to-day ethical decision-making.
This gap is filled by mentor support.
P16 advocated for significant changes:

\textit{P16: ``We need ideally partially blind distributed ethical review that also accounts for things like some things are plain wrong, like everyone globally would agree that it is wrong. [...] But then there's also like these bigger grey areas where what is considered right or wrong deviates."}

Researchers often work without any official ethical guidelines but rather they put together ethical ideas from various sources.
These personal methods might not be the same for everyone as some rely on \textit{``common sense''} (P6) and others learn about ethics only when they need to (P2).
This informal approach allows the researchers adjust to different situations, but it also means that ethical considerations might be missed or handled differently each time.

\subsubsection{Ethical Reporting in Publications} 
\label{publication requirements} 

Internet measurement conferences have different ethics requirements and researchers have different ways of handling these.
 
\paragraph{\textbf{Variation Across Venues}}
 
\textit{``Different venues will have different ethics standards''} (P1).
One venue has a \textit{``high standard of ethics and another venue that should have a really high standard of ethics also don't''} (P1).
Even venues with high ethics standards \textit{``accept papers that should not be accepted because they clearly have ethical considerations that were not taken into account''} (P1).
This challenge is further compounded:

\textit{P1: ``ethical boards [...] have different standards, but also different conferences are going to have different PCs\footnote{Program Committees} and these members are going to have different standards. So, since we don't have a minimal bar for ethical standards, because it's difficult not to say it's an easy task, ethics change over time.''}

Hence, while ethics sections are becoming a requirement, it can be \textit{``performative''} (P16), \textit{``not necessarily enforced at the end''} (P15) and the content might not be rigorously checked (P14), also noted in Section~\ref{uncertainethics}.

\paragraph{\textbf{Is a One-Line Statement Enough?}}
 
Our analysis of the three conferences\footnote{IMC, PAM and TMA} showed that the amount of papers that actually discussed the ethics and the papers that did not discuss it at all or just gave a one-liner varied.

\textit{P14: ``I think often something slips through where people just write a statement, a one-liner, just where it does not raise ethical concerns. And it does raise ethical concerns, but no one really noticed or cared."}

P16 explained why short statements might be okay. 

\textit{P16: ``Some people just say... ``Our work doesn't raise ethical considerations'' [...] I think IMC even has a pre-formulated one-line sentence for that. I mean if you put like a couple of computers in a rack and do some in-house measurements between those computers do you have to think about ethics? Do you have to like elaborately write a paragraph? What would you write in that paragraph?"}

\textit{P16: ``Then again if you do performance measurements of switch throughput [...] for switches that are supposed to be put into autonomous fighting drones - different thing."}

The challenge lies, therefore, in distinguishing between research that genuinely raises minimal ethical concerns and research where superficial statements mask significant ethical issues.
This responsibility falls on the conference reviewers who need to be well-informed because \textit{``the person that needs to impose most of these rules are the reviewers in the venues''} (P1) and \textit{``the quality of the reviewers and the understanding of the reviewers are the key elements''} (P1).
This goes beyond peer-review because \textit{``to have higher standards of ethics, you need to better inform reviewers. To have better informed reviewers, you need to have PhD programs that inform the PhD students. To have better PhD programs that inform the PhD students, you need to have universities that discuss between each other ethical standards''} (P1).
Opinions on a one-line statement vary:

\textit{P15: ``I think if it's only a one-liner, that's not okay. So at least some discussion, [...] what they actually considered, what they thought of, why it's not necessary to have ethical measures, like that is a consideration as well that should be discussed to some degree."}

However, depending on the type of Internet measurement, ethics considerations can change drastically.
For instance, while traffic analysis \textit{``has an ethics component''} and papers often include \textit{``a four or five sentence ethics concern section''}, \textit{``censorship [research] is more strict towards ethics''} involving \textit{``a lot more work''} and a more detailed ethics section.

\subsubsection{Evolution of Ethics}
\textit{``Ethics change over time''} (P1).

\textit{P1: ``When I started doing research we had some ethical considerations but we had way more flexibility in what we were doing.”}

Previously, ethics \textit{``were too loose''} (P11) and rarely discussed in published Internet measurement research.
Now \textit{``at least the main conferences or the main research group in this area has settled on some unique or some collaborative measures that are implemented''} (P14).
Although ethics considerations might be limiting, \textit{``it limits the research in a good way. Because a lot of the questions that you don’t think about the ethics implications, you are probably doing something wrong''} (P1).
Participants noted how formal rules have changed over time and how cultural practices have been passed down through different research groups.

\textit{P1: ``Over time, I saw topics that were impossible to do research around because of ethical considerations, became possible and now they are quite popular."}

Ethical boundaries are not simply becoming more restrictive over time, but are evolving in complex ways that both constrain and enable different types of research, reflecting broader cultural shifts in how the research community understands good practice.
However, as discussed in Sections~ \ref{difficulttransparent}, \ref{uncertainethics} and~\ref{publication requirements}, formal requirements might not substitute for genuine ethical consideration. 

\section{Discussion} \label{Discussion}
Active measurements create unpredictable and often immediate ethical challenges for researchers, requiring
sophisticated technical coping mechanisms, e.g. mandatory lab testing before live scanning, carefully calibrated rate-limits, and detailed transparency mechanisms. 
Controlled lab environments and small-scale Internet tests before full deployment represents proactive risk management.
Legal ambiguity also looms large; 
researchers received complaints from operators and even attracted attention from law enforcement. 
Such work can be difficult for outsiders to distinguish from malicious activity, highlighting the gray legal boundaries of active measurements. 
These issues become more complex when it comes to censorship studies as it often involves activity that is unlawful in the countries they study, raising concerns about participant safety, researcher liability, and methodological legitimacy. 
Researchers rely on detailed risk assessment protocols, including careful evaluation of research benefits against potential harm to participants and researchers,
sometimes leading to research abandonment.

Passive measurement researchers face fewer and less immediate ethical challenges. 
While their research includes working with publicly available data, 
hidden identifiers in public data sets and data inference were the two most common challenges.
They used anonymization techniques and access control mechanisms to address these. 
Data that appears technical can still reveal sensitive information, especially when analyzed and combined, something not accounted for in established processes.
Moreover, researchers with commercial datasets are bound be confidentiality.
These findings suggest that passive measurements may create a false sense of ethical security. The indirect nature of potential harms makes them harder to anticipate and prevent which may lead to more widespread but less visible ethical violations. 

\subsection{The Hidden Labor} 
\label{The Invisible Labor}
Internet measurement researchers perform additional--often invisible--work, not discussed in existing literature.
This \textit{ethical labor} is a hidden cost of responsible research practices.

\paragraph{\textbf{Continuous Monitoring and Availability}}
Active measurement researchers monitor the scans vigilantly by checking for and dealing with abuse emails and complaints even on weekends, holidays, and outside regular work hours.
They avoid starting scans on the weekends because they might not be able to respond quickly.
This can cause stress, and requires resources beyond institutional support.

\paragraph{\textbf{Engaging with Operators and their Frustrations}}
Researchers often have to thoroughly explain their research methods (including ethical guidelines and research benefits) to network operators, who may have concerns about legitimacy. 
However, operators may become frustrated and abusive.
Beyond technical prowess, such work demands emotional labor, communication, and conflict resolution skills that the researchers might not be trained for.

\paragraph{\textbf{Community Responsibility}}
Researchers feel responsible for maintaining honesty and trustworthiness in the Internet measurement community. 
In addition to finding ways to raise the research standards of the whole community, they regularly consider the reputation of the community as well. 
Many worry that a single mistake could cast doubt on the research field as a whole.

\subsection{Insights and Recommendations}
    \paragraph{\textbf{Recognize and support the hidden labor}} The informal work outlined in Section~\ref{The Invisible Labor} constitutes the hidden support infrastructure upholding responsible and ethical Internet measurements. Such work involves dedication and care, and must be recognized and supported as such.
    \paragraph{ \textbf{The craft of ethics}} Amidst a lack of frameworks and reliance on mentors - the practice of ethics emerges as a type of \textit{craft knowledge}.
    It is not learned via formal methods but through trial and error, experience, community norms, and generational knowledge transfer.
    As such, we need better ways to capture and disseminate this wisdom, both within research groups and the larger community.

\subsection{Limitations and Future Work}
Researchers' experiences from one Internet measurement group do not represent the experiences of the Internet measurement community as a whole.
Our study can be used as a springboard for future studies exploring researchers' experiences from different research groups and jurisdictions.

Our work has limitations common for qualitative research.
We mitigate participant subjectivity by relying on methodological rigor, e.g. creating a detailed interview guide and soliciting actual examples from participants in addition to their opinions.
We reduce researcher subjectivity through reflexivity and team discussions to interpret the findings.
\section{Conclusion} 
\label{conclusion}

We engaged in conversations with sixteen Internet measurement researchers from one research group.
We found that instead of following any established and available guidelines, they use a mix of their own judgment, rely on their supervisors and mentors' input, knowledge shared in the published papers, and sometimes even follow the guidelines of the tools that they use. 
They think carefully about how to conduct ethically sound research while making sure they do not cause any harm to the system or the people they are studying. 
Our findings reveal that ethical decision-making in this field cannot be separated from technical expertise, underscoring the importance of having reviewers with sufficient domain knowledge to evaluate the ethical dimensions of studies. 
We also found that coping strategies are mostly passed along informally through mentorship, yet these lessons could benefit the broader community if captured and shared more systematically. 
Finally, the invisible labor required to maintain trust—educating operators, monitoring systems during off-hours, and ensuring that no harm occurs—remains crucial and must be supported.

\section{Ethics Considerations}
The project description, methodology, recruitment strategy, data collection and storage procedures, and informed consent form (see Appendix \ref{Consent Form}) were submitted for review to the ethics committee, for which approval was received.
Personally Identifiable Information (PII), such as audio recordings, was deleted after transcription and de-identification.
Similarly, consent forms containing participant names were also deleted after project completion.
All personal and sensitive data were stored and processed on internal servers, as such no personal data was shared with third parties.
Participants were provided details about data collection, data processing and storage, along with a project summary (see Appendix \ref{project desc}), via informed consent forms.
We also did not connect participant IDs (e.g. P1, P2) to participant information in Table~\ref{tab:comprehensive_background} to further reduce the risk of deanonymization.


\bibliographystyle{IEEEtran}
\bibliography{ref}

@inproceedings{Fiebig2024,
author = {Fiebig, Tobias},
title = {Crisis, Ethics, Reliability \& a measurement.network: Reflections on Active Network Measurements in Academia},
year = {2023},
isbn = {9798400702747},
publisher = {Association for Computing Machinery},
address = {New York, NY, USA},
url = {https://doi.org/10.1145/3606464.3606483},
doi = {10.1145/3606464.3606483},
booktitle = {Proceedings of the 2023 Applied Networking Research Workshop},
pages = {44–50},
numpages = {7},
location = {San Francisco, CA, USA},
series = {ANRW '23}
}

@inproceedings{10.1145/2793013.2793021,
author = {Crandall, Jedidiah R. and Crete-Nishihata, Masashi and Knockel, Jeffrey},
title = {Forgive Us our SYNs: Technical and Ethical Considerations for Measuring Internet Filtering},
year = {2015},
isbn = {9781450335416},
publisher = {Association for Computing Machinery},
address = {New York, NY, USA},
url = {https://doi.org/10.1145/2793013.2793021},
doi = {10.1145/2793013.2793021},
booktitle = {Proceedings of the 2015 ACM SIGCOMM Workshop on Ethics in Networked Systems Research},
pages = {3},
numpages = {1},
location = {London, United Kingdom},
series = {NS Ethics '15}
}

@inproceedings{Partridge2016,
  author = {Partridge, Craig and Allman, Mark},
  title = {Ethical considerations in network measurement papers},
  year = {2016},
  issue_date = {October 2016},
  publisher = {Association for Computing Machinery},
  address = {New York, NY, USA},
  volume = {59},
  number = {10},
  issn = {0001-0782},
  url = {https://doi.org/10.1145/2896816},
  doi = {10.1145/2896816},
  journal = {Commun. ACM},
  month = sep,
  pages = {58–64},
  numpages = {7}
}

@inproceedings{Kohno2023,
  author = {Tadayoshi Kohno and Yasemin Acar and Wulf Loh},
  title = {Ethical Frameworks and Computer Security Trolley Problems: Foundations for Conversations},
  booktitle = {32nd USENIX Security Symposium (USENIX Security 23)},
  year = {2023},
  isbn = {978-1-939133-37-3},
  address = {Anaheim, CA},
  pages = {5145--5162},
  url = {https://www.usenix.org/conference/usenixsecurity23/presentation/kohno},
  publisher = {USENIX Association},
  month = aug
}

@inproceedings {DScope2023,
author = {Eric Pauley and Paul Barford and Patrick McDaniel},
title = {{DScope}: A {Cloud-Native} Internet Telescope},
booktitle = {32nd USENIX Security Symposium (USENIX Security 23)},
year = {2023},
isbn = {978-1-939133-37-3},
address = {Anaheim, CA},
pages = {5989--6006},
url = {https://www.usenix.org/conference/usenixsecurity23/presentation/pauley},
publisher = {USENIX Association},
month = aug
}

@inproceedings{Mazel2017,
  author={Mazel, Johan and Fontugne, Romain and Fukuda, Kensuke},
  booktitle={2017 Network Traffic Measurement and Analysis Conference (TMA)}, 
  title={Profiling internet scanners: Spatiotemporal structures and measurement ethics}, 
  year={2017},
  volume={},
  number={},
  pages={1-9},
  doi={10.23919/TMA.2017.8002909}
}

@article{Menlo2012,
  title={The menlo report: Ethical principles guiding information and communication technology research},
  author={Kenneally, Erin and Dittrich, David},
  journal={Available at SSRN 2445102},
  year={2012},
doi = {10.2139/ssrn.2445102}
}

@misc{belmont1979,
  author = {{National Commission for the Protection of Human Subjects of Biomedical and Behavioral Research}},
  title       = {The Belmont Report: Ethical Principles and Guidelines for the Protection of Human Subjects of Research},
  year = {1979},
  month = apr,
  publisher = {U.S. Department of Health \& Human Services},
  url = {https://www.hhs.gov/ohrp/regulations-and-policy/belmont-report/read-the-belmont-report/index.html},
  note = {Accessed: 2024-11-01},
}

@article{Ciprian2018,
  title={Ethics, Products, Top Lists-and their Use at Internet Measurement Conferences},
  author={Ciprian, Luca},
  journal={Network},
  volume={23},
  year={2018}
}

@misc{IMC2024,
  author = {},
  title = {{IMC 2024 Submission Instructions}},
  year = {2024},
  url = {https://conferences.sigcomm.org/imc/2024/submission-instructions/},
  note = {Accessed: 2024-12-11},
 
}

@inproceedings{Dirksen2024,
  author = {Dirksen, Alexandra and Giessler, Sebastian and Erz, Hendrik and Johns, Martin and Fiebig, Tobias},
  title = {Don't Patch the Researcher, Patch the Game: A Systematic Approach for Responsible Research via Federated Ethics Boards},
  booktitle = {Proceedings of the 2023 ACM SIGSAC Conference on Computer and Communications Security},
  year = {2024},
  pages = {126-141}, 
  publisher = {ACM},
  doi = {10.1145/3703465.3703475},
  url = {https://doi.org/10.1145/3703465.3703475}
}

@misc{PAM2024,
  author = {},
  title = {{Passive and Active Measurement Conference 2024 Call for Papers}},
  year = {2024},
  url = {https://pam2024.cs.northwestern.edu/cfp/},
  note = {Accessed: 2024-12-11},  
}

@online{TMA2024,
  author = {},
  title = {{TMA 2024 Submission Instructions}},
  year = {2024},
  url = {https://tma.ifip.org/2024/submission-instructions/},
  note = {Accessed: 2024-12-11},
}

@INPROCEEDINGS{Zhang2022,
  author={Zhang, Yiming and Liu, Mingxuan and Zhang, Mingming and Lu, Chaoyi and Duan, Haixin},
  booktitle={2022 IEEE European Symposium on Security and Privacy Workshops (EuroS\&PW)}, 
  title={Ethics in Security Research: Visions, Reality, and Paths Forward}, 
  year={2022},
  volume={},
  number={},
  pages={538-545},
  doi={10.1109/EuroSPW55150.2022.00064}
}

@inproceedings{pauley2023,
  address   = {San Diego, CA},
  title     = {Understanding the Ethical Frameworks of Internet Measurement Studies},
  booktitle = {The 2nd International Workshop on Ethics in Computer Security (EthiCS 2023)},
  author    = {Pauley, Eric and McDaniel, Patrick},
  month     = feb,
  year      = {2023},
  doi       = {10.14722/ethics.2023.239547}
}

@INPROCEEDINGS {Florian,
author = { Hantke, Florian and Roth, Sebastian and Mrowczynski, Rafael and Utz, Christine and Stock, Ben },
booktitle = { 2024 IEEE Symposium on Security and Privacy (SP) },
title = {{ Where Are the Red Lines? Towards Ethical Server-Side Scans in Security and Privacy Research }},
year = {2024},
volume = {},
ISSN = {},
pages = {4405-4423},
doi = {10.1109/SP54263.2024.00104},
url = {https://doi.ieeecomputersociety.org/10.1109/SP54263.2024.00104},
publisher = {IEEE Computer Society},
address = {Los Alamitos, CA, USA},
month =May}

@inproceedings{Huz2015MTurk,
author = {Huz, Gokay and Bauer, Steven and claffy, kc and Beverly, Robert},
year = {2015},
month = {08},
pages = {27-32},
booktitle ={},
title = {Experience in using MTurk for Network Measurement},
doi = {10.1145/2787394.2787399}
}

@article{Finn2023Menlo,
  title={Ethics governance development: The case of the Menlo Report},
  author={Finn, Megan and Shilton, Katie},
  journal={Social Studies of Science},
  volume={53},
  number={3},
  pages={315--340},
  year={2023},
  publisher={SAGE Publications Sage UK: London, England}
}

@inproceedings{Ethics2017,
author = {Van der Ham, Jeroen},
year = {2017},
month = {05},
pages = {247-251},
booktitle ={},
title = {Ethics and Internet Measurements},
doi = {10.1109/SPW.2017.17}
}

@article{soe2023ethics,
  title={The ethics of sharing: privacy, data, and common goods},
  author={S{\o}e, Sille Obelitz and Mai, Jens-Erik},
  journal={Digital Society},
  volume={2},
  number={2},
  pages={28},
  year={2023},
  publisher={Springer}
}

@inproceedings {ZMAP,
author = {Zakir Durumeric and Eric Wustrow and J. Alex Halderman},
title = {{ZMap}: Fast Internet-wide Scanning and Its Security Applications},
booktitle = {22nd USENIX Security Symposium (USENIX Security 13)},
year = {2013},
isbn = {978-1-931971-03-4},
address = {Washington, D.C.},
pages = {605--620},
url = {https://www.usenix.org/conference/usenixsecurity13/technical-sessions/paper/durumeric},
publisher = {USENIX Association},
month = aug
}

@article{Encore,
author = {Burnett, Sam and Feamster, Nick},
title = {Encore: Lightweight Measurement of Web Censorship with Cross-Origin Requests},
year = {2015},
issue_date = {October 2015},
publisher = {Association for Computing Machinery},
address = {New York, NY, USA},
volume = {45},
number = {4},
issn = {0146-4833},
url = {https://doi.org/10.1145/2829988.2787485},
doi = {10.1145/2829988.2787485},
journal = {SIGCOMM Comput. Commun. Rev.},
month = aug,
pages = {653–667},
numpages = {15}
}

@article{Braun08082019,
author = {Virginia Braun and Victoria Clarke},
title = {Reflecting on reflexive thematic analysis},
journal = {Qualitative Research in Sport, Exercise and Health},
volume = {11},
number = {4},
pages = {589--597},
year = {2019},
publisher = {Routledge},
doi = {10.1080/2159676X.2019.1628806},
URL ={https://doi.org/10.1080/2159676X.2019.1628806},
eprint = { https://doi.org/10.1080/2159676X.2019.1628806}}

@article{Braun01012006,
author = {Virginia Braun and Victoria Clarke},
title = {Using thematic analysis in psychology},
journal = {Qualitative Research in Psychology},
volume = {3},
number = {2},
pages = {77-101},
year = {2006},
publisher = {Routledge},
doi = {10.1191/1478088706qp063oa},
URL ={https://doi.org/10.1191/1478088706qp063oa},
eprint = { https://doi.org/10.1191/1478088706qp063oa}}

@article{wendzel2025survey,
  author = {Steffen Wendzel and
Simon Volpert and Sebastian Zillien and Julia Lenz and Philip R{\"{u}}nz and Luca Caviglione},
  title = {A Survey of Internet Censorship and its Measurement: Methodology, Trends, and Challenges},
  journal = {CoRR},
  volume  = {abs/2502.14945},
  year    = {2025},
  url     ={https://doi.org/10.48550/arXiv.2502.14945},
  doi          = {10.48550/ARXIV.2502.14945},
  eprinttype    = {arXiv},
  eprint       = {2502.14945},
  bibsource    = {dblp computer science bibliography, https://dblp.org}
}

@article{kaur2022needed,
author = {Kaur, Mannat and Parkin, Simon and Janssen, Marijn and Fiebig, Tobias},
title = {"I needed to solve their overwhelmness": How System Administration Work was Affected by COVID-19},
year = {2022},
issue_date = {November 2022},
publisher = {Association for Computing Machinery},
address = {New York, NY, USA},
volume = {6},
number = {CSCW2},
url = {https://doi.org/10.1145/3555115},
doi = {10.1145/3555115},
journal = {Proc. ACM Hum.-Comput. Interact.},
month = nov,
articleno = {390},
numpages = {30}
}

@article{Kaur23,
author = {Kaur, Mannat and Sri Ramulu, Harshini and Acar, Yasemin and Fiebig, Tobias},
title = {"Oh yes! over-preparing for meetings is my jam:)": The Gendered Experiences of System Administrators},
year = {2023},
issue_date = {April 2023},
publisher = {Association for Computing Machinery},
address = {New York, NY, USA},
volume = {7},
number = {CSCW1},
url = {https://doi.org/10.1145/3579617},
doi = {10.1145/3579617},
journal = {Proc. ACM Hum.-Comput. Interact.},
month = apr,
articleno = {141},
numpages = {38}
}

@InCollection{sep-ethics-internet-research,
	author       =	{Buchanan, Elizabeth A. and Zimmer, Michael},
	title        =	{{Internet Research Ethics}},
	booktitle    =	{The {Stanford} Encyclopedia of Philosophy},
	editor       =	{Edward N. Zalta and Uri Nodelman},
	howpublished =	{\url{https://plato.stanford.edu/archives/fall2025/entries/ethics-internet-research/}},
	year         =	{2025},
	edition      =	{{F}all 2025},
	publisher    =	{Metaphysics Research Lab, Stanford University}
}

@article{Braun03072021,
author = {Virginia Braun and Victoria Clarke},
title = {One size fits all? What counts as quality practice in (reflexive) thematic analysis?},
journal = {Qualitative Research in Psychology},
volume = {18},
number = {3},
pages = {328--352},
year = {2021},
publisher = {Routledge},
doi = {10.1080/14780887.2020.1769238},
URL ={https://doi.org/10.1080/14780887.2020.1769238},
eprint ={https://doi.org/10.1080/14780887.2020.1769238}}

\appendices
\section{Consent Form} \label{Consent Form}
\subsection{Taking part in the study}
\begin{enumerate}
    \item I have read and understood the study information dated, or it has been read to me. I have been able to ask questions about the study and my questions have been answered to my satisfaction.
    \item I consent voluntarily to be a participant in this study and understand that I can refuse to answer questions and I can withdraw from the study at any time, without having to give a reason.
    \item I understand that taking part in the study involves:
    \begin{itemize}
        \item Participating in an audio/video recorded interview via a self-hosted platform (BBB\footnote{Big Blue Button}), which will take approximately 45-60 minutes.
        \item My responses will be transcribed and anonymized for analysis, and the recordings will be destroyed after transcription.
    \end{itemize}
    \item I understand that I will not be compensated for my participation.
    \item I understand that taking part in the study involves collecting specific personally identifiable information (PII) and associated personally identifiable research data (PIRD) such as:
    \begin{itemize}
        \item Name
        \item Email Address
        \item Professional Role and Field of Expertise
        \item Job location and Years of Experience
        \item Gender and Ethnicity
    \end{itemize}
    \item I understand that the following steps will be taken to minimize the threat of a data breach and protect my identity:
    \begin{itemize}
        \item Data will be anonymized before analysis.
        \item Audio recordings will be destroyed after transcription.
        \item All data collected will only be stored internally on protected servers and access to data will be restricted to authorized research team members only.
    \end{itemize}
    \item I understand that personal information about me that could identify me will not be shared beyond the research team.
    \item I understand that the (identifiable) personal data I provide will be destroyed after the completion of the study.
    \item  I understand that after the research study the de-identified information I provide will be used for:
    \begin{itemize}
        \item Academic publications, including peer-reviewed journal articles, conference papers, and other scholarly outputs.
        \item Presentations at academic conferences, workshops, and seminars to disseminate the findings to the research community.
        \item Policy development or recommendations, where applicable, particularly in the areas of ethical practices in network measurement research.
        \item Development of guidelines or best practices for the research community, based on the ethical challenges and solutions identified in the study.
        \item The data may also be used in future research studies and publications to advance knowledge in the field of network measurement research ethics.
    \end{itemize}
    \item I agree that my responses, views or other input can be quoted anonymously in research outputs

\end{enumerate}

\subsection{Project Description in the Consent Form}
\label{project desc}
\paragraph{Motivation:} Network measurement research plays a pivotal role in understanding and improving internet infrastructure and performance. The ethical implications of network measurement research have become increasingly evident as data collection scales up and research methods evolve. While several studies have explored these ethical challenges, particularly through surveys and interviews with network operators, legal experts, and ethics committee members, there remains a gap in understanding how ethical decisions are made in practice by network measurement researchers themselves. Existing studies tend to focus on aggregated data from external stakeholders rather than the actual decision-making processes of researchers conducting network measurement studies. Our study aims to fill this gap by interviewing network measurement researchers directly.

\paragraph{Our Study:} This study seeks to explore how network measurement researchers approach the ethical aspects of their research. Through in-depth interviews, we aim to gather insights into the research practices, ethical frameworks and guidelines that shape their research decisions. By focusing on the personal experiences of researchers, our goal is to provide a deeper, more comprehensive understanding of how ethical considerations are integrated into network measurement research, going beyond the surveys and aggregated data commonly used in past studies.

\section{Interview Structure } \label{Interview Structure}
We started off with general question (1) about their research scope and then depending on the answers of the participants we asked the next questions. The interview questions were not in one order for all the participants. We followed the natural flow of the conversation.

\begin{enumerate}
    \item Can you briefly describe your work in network measurement research?
    \begin{enumerate}
        \item What are the primary goals of your research?
        \item What challenges do you commonly face in this field?
        \item What steps do you take when setting up a network measurement study?
    \end{enumerate}
    \item Are ethical considerations a part of your network measurement research?\newline
    \textbf{IF YES:}
    \begin{enumerate}
        \item In what ways do ethics influence your research?
        \begin{enumerate}
            \item Do you perform an ethics review prior to conducting your research?\newline
            \textbf{IF YES:}
            \begin{enumerate}
                \item What does that process look like?
            \end{enumerate}
            \begin{enumerate}
                \item Do you take approval for the Ethics Review board? 
            \end{enumerate} 
            
            \textbf{IF NO:}
            \begin{enumerate}
                \item Do you collect any personal data whilst conducting your research?                
            \end{enumerate}
            \textbf{IF YES:}
            \begin{enumerate}
                \item What challenges do you face in getting approval from Institutional Review Boards (IRBs) for your network measurement studies?
            \end{enumerate}
            \item Do you refer to ethical guidelines/frameworks?
            \textbf{IF YES: }
            \begin{enumerate}
                \item Which ones?
            \end{enumerate}
            \item Can you describe a specific instance where you encountered ethical challenges during your network measurement research? \newline
            \textbf{Follow-up}
            \begin{enumerate}
                \item How did you address these challenges?
            \end{enumerate}
        \end{enumerate}
    \end{enumerate}
    \textbf{IF NO: }
    \begin{enumerate}
        \item Why do you think ethics are not a major factor in your work?
    \end{enumerate}
    \item Is informed consent required in your network measurement studies? \newline
    \textbf{If Yes (Consent is required):}
    \begin{enumerate}
        \item How is Informed consent obtained?
        \begin{enumerate}
            \item When is it required, and when might it be impractical?
            \item How do you ensure users understand the implications of your research?
        \end{enumerate}

    \textbf{If No (Consent is not required or impractical):}
    
    \item Why is informed consent not required?
    \begin{enumerate}
        \item What factors make informed consent difficult or impractical in network measurement studies?
        \item If obtaining informed consent is not feasible or not required, what measures do you take to ensure transparency and accountability in your research?
    \end{enumerate}
\end{enumerate}
\item Do you come across privacy concerns?\newline
\textbf{IF YES:}
\begin{enumerate}
    \item How do you handle privacy concerns when collecting network measurement data?
    \item Do you implement any safeguards to protect user privacy?
    \item Have you faced situations where user data was inadvertently exposed?
    \item How do you mitigate potential privacy risks after data has already been collected?
\end{enumerate}
\item Do you think your research can lead to unintended harm to individuals and networks?\newline
\textbf{IF YES:}
\begin{enumerate}
    \item How do you handle that?
\end{enumerate}
\textbf{IF NO:}
\begin{enumerate}
    \item Have you ever encountered a situation where network measurement caused disruptions or security concerns? \newline
    \textbf{Follow-up:}
    \begin{enumerate}
        \item How did you handle it?
    \end{enumerate}
    \item How do you balance the need for innovation in network measurement research with the potential ethical risks to users or networks?
\end{enumerate}
    \item What do you think about the current ethical standards of network measurement research?
    \begin{enumerate}
        \item Do you think they are sufficient to address the associated risks?\newline
        \textbf{If Yes (They are sufficient):}
        \begin{enumerate}
            \item What aspects of the current standards do you find effective?
            \item How do these standards support you in making ethical decisions?
        \end{enumerate}
        \textbf{If No (They need improvement):}
        \begin{enumerate}
            \item What gaps or limitations do you see in the current ethical guidelines?
            \item What changes would you suggest, to improve ethical compliance in network measurement research?
        \end{enumerate}
    \end{enumerate}
    \item Do you have any recommendations for other researchers regarding the ethical aspects of network measurement work?
    \item Do you see any difference between how you dealt with ethics in the past versus today?
    \begin{enumerate}
        \item Have you ethical considerations changed over the years?
    \end{enumerate}
\end{enumerate}

\section{Codebook} 
\label{subsec:codebook}
We present an overview of the study’s codebook, which comprises 1,226 codes organized into thematic categories that cover research practices, ethical considerations, challenges, and recommendations in~\autoref{tab:Codebook}. 
It highlights the range and distribution of topics addressed, from privacy concerns and risk mitigation to transparency, accountability, and evolving ethical standards.

\begin{table}[ht!]
\centering
\caption{\textbf{Overview of the codebook (1,226 codes)}}
\label{tab:Codebook}
\begin{tabular}{l c}
\toprule
\textbf{Code} & \textbf{\#} \\
\midrule
Current Research Focus & 32 \\
Ethical Considerations & 136 \\
Challenges & 80 \\
Balancing Innovation vs Ethical Risks & 68 \\
Privacy Concerns & 54 \\
Unintended Harm & 41 \\
Ethics Review & 51 \\
Past Ethics vs Now & 25 \\
Transparency and Accountability & 47 \\
Informed Consent & 13 \\
Study Setup Steps \& Challenges & 43 \\
Current Ethical Standards & 98 \\
Ethical Challenges & 16 \\
Ethical Guidelines & 56 \\
How to Handle the Privacy Concerns & 31 \\
Risk Mitigation Strategies & 16 \\
IRB and Review Process & 21 \\
Recommendations & 94 \\
Improving Current Ethical Standards & 29 \\
Other Themes & 275 \\
\midrule
\textbf{Total} & \textbf{1,226} \\
\bottomrule
\end{tabular}
\end{table}

\end{document}